\documentclass[%
 aip,
 amsmath,amssymb,
preprint,%
]{revtex4-1}

\usepackage{graphicx}
\usepackage{dcolumn}
\usepackage{bm}

\usepackage[utf8]{inputenc}
\usepackage[T1]{fontenc}
\usepackage{mathptmx}
\usepackage{etoolbox}
\usepackage{subfigure}
\makeatletter
\def\@email#1#2{%
 \endgroup
 \patchcmd{\titleblock@produce}
  {\frontmatter@RRAPformat}
  {\frontmatter@RRAPformat{\produce@RRAP{*#1\href{mailto:#2}{#2}}}\frontmatter@RRAPformat}
  {}{}
}%
\makeatother
\begin{document}

\preprint{DRAFT: Structural Dynamics}

 \title[]{A machine learning photon detection algorithm for coherent X-ray ultrafast fluctuation analysis}

\author{Sathya R. Chitturi*}
\email{chitturi@stanford.edu}
\affiliation{\mbox{Department of Materials Science and Engineering, Stanford University, Stanford, USA}}%
\affiliation{SLAC National Accelerator Laboratory, Menlo Park, USA}%
\affiliation{\mbox{Stanford Institute for Materials and Energy Sciences, Stanford, USA}}%

\author{Nicolas G. Burdet}%
\affiliation{SLAC National Accelerator Laboratory, Menlo Park, USA}%

\author{Youssef Nashed}%
\affiliation{SLAC National Accelerator Laboratory, Menlo Park, USA}%

\author{Daniel Ratner}%
\affiliation{SLAC National Accelerator Laboratory, Menlo Park, USA}%

\author{Aashwin Mishra}%
\affiliation{SLAC National Accelerator Laboratory, Menlo Park, USA}%

\author{TJ Lane}%
\affiliation{ 
Deutsches Elektronen-Synchrotron, Hamburg, Germany}%

\author{Matthew Seaberg}%
\affiliation{SLAC National Accelerator Laboratory, Menlo Park, USA}%

\author{Vincent Esposito}%
\affiliation{SLAC National Accelerator Laboratory, Menlo Park, USA}%

\author{Chun H. Yoon}%
\affiliation{SLAC National Accelerator Laboratory, Menlo Park, USA}%

\author{Mike Dunne}%
\affiliation{SLAC National Accelerator Laboratory, Menlo Park, USA}%

\author{Joshua J. Turner}
\affiliation{SLAC National Accelerator Laboratory, Menlo Park, USA}%
\affiliation{\mbox{Stanford Institute for Materials and Energy Sciences, Stanford, USA}}%

\date{\today}

\begin{abstract}
X-ray free electron laser (XFEL) experiments have brought unique capabilities and opened new directions in research, such as creating new states of matter or directly measuring atomic motion. One such area is the ability to use finely spaced sets of coherent x-ray pulses to be compared after scattering from a dynamic system at different times. This enables the study of fluctuations in many-body quantum systems at the level of the ultrafast pulse durations, but this method has been limited to a select number of examples and required complex and advanced analytical tools. By applying a new methodology to this problem, we have made qualitative advances in three separate areas that will likely also find application to new fields. As compared to the `droplet-type' models which typically are used to estimate the photon distributions on pixelated detectors to obtain the coherent X-ray speckle patterns, our algorithm pipeline achieves an order of magnitude speedup on CPU hardware and two orders of magnitude improvement on GPU hardware. We also find that it retains accuracy in low-contrast conditions, which is the typical regime for many experiments in structural dynamics. Finally, it can predict photon distributions in high average-intensity applications, a regime which up until now, has not been accessible. Our AI-assisted algorithm will enable a wider adoption of x-ray coherence spectroscopies, by both automating previously challenging analyses and enabling new experiments that were not otherwise feasible without the developments described in this work.
\end{abstract}

\maketitle

\section{\label{sec:Introduction}Introduction}

The construction and operation of X-ray free electron lasers (XFELs) \cite{decking-natphot-2020, prat-2020-natphys, kang-natphot-2017, ishikawa-2012-natphot, Emma-2010-NatPho} has enabled a great leap towards deeper understanding of a diverse area of scientific research areas \cite{bostedt2016linac}, including planetary science \cite{Vinko-2012-Nature}, astrophysics \cite{Bernitt-2012-Nature}, medicine \cite{Redecke-2013-Science} and molecular chemistry \cite{Wernet-2015-Nature}. With the unprecedented brightness, short pulse duration, and x-ray wavelengths, new states of matter can be created and studied \cite{lee-2021-prl}, while dynamics can be monitored, and now controlled, on ultrafast timescales \cite{wandel-science-2022}.

With the start of high repetition rate next-generation light sources, methods which have so far been challenging will become feasible, such as resonant inelastic x-ray scattering at high time- and spectra- resolution \cite{Ament-2011-RMP} and x-ray photoemission spectroscopy \cite{Siefermann-2014}. One such example is X-ray photon correlation spectroscopy (XPCS) \cite{sutton-2008-crp,Shpyrko-Nature-2007}, which uses the spatial coherence of the X-ray beam to produce a scattering `fingerprint' of the sample. This fingerprint, or speckle pattern, can be correlated in time to directly observe equilibrium dynamics of a given system. This information of the thermal fluctuations can be related back the energetics and the interactions in the system. This is typically measured by calculating the intensity-intensity autocorrelation function and extracting the intermediate scattering function $S(q,t)$ (Equation \ref{eqn:Siegert}),

\begin{equation}
g_2(q,t) = 1+A{S(q,t)}^2
\label{eqn:Siegert}
\end{equation}
which allows the time correlation to be related back to the physical properties of the system being studied.

Another benefit of these new machines is in their ability to produce finely spaced X-ray pulses with controllable delay, using X-ray optics \cite{sun-ol-2019} or special modes of the accelerator \cite{decker-2022-SciRep}. These pulses enable studies of spontaneous fluctuations at orders of magnitude faster timescales than what is possible using XPCS at x-ray synchrotron facilities, with one key area of application being emergent phenomena in quantum materials. We refer to this multi-pulse adding technique here as X-ray photon fluctuation spectroscopy (XPFS) \cite{Shen2021}. This is a unique tool which differs from traditional pump-probe spectroscopy which detects the relaxation from a non-equilibrium state, by instituting more of a `probe-probe' method, where fluctuations in the equilibrium state can be measured directly by comparing how the system changes between probe pulses. 
Here, one adds the pulses which are too close together in time to be read out by the detector \cite{Gutt-2009-OptExp} and uses statistics of the coherent speckle \cite{goodman-2007-book} to compute the fluctuation spectra using the contrast \cite{decaro-jsr-2013,bandyopadhyay-2005-rsi}, i.e. the fast dynamical information of the system can be distinguished by studying single photon fluctuations. Even with the massive amount of photons per pulse, three things typically result in a single photon detection process: the decrease of intensity after the scattering process on a single pulse basis, the short pulse duration, and the sometimes reduced intensity required to ensure excitations are not produced in the sample.

In principle, if the discrete distribution of photon counts over the detector can be accurately measured and enough samples averaged, it is possible to  determine the dynamical evolution the sample by computing the speckle contrast $C\left(q,t\right)$ as a function of delay-time $t$ and momentum transfer $q$. The contrast is obtained by fitting a negative binomial distribution parameterized by $M \equiv M(q,t)$ = $\frac{1}{C^2\left(q,t\right)}$ and the average number of photons per pixel $\bar{k}$ \cite{Hruszkewycz-2012-PRL} -- i.e. $P\left(k; \bar{k}, M\right)$:


\begin{equation}
P(k, \bar{k}, M)=\frac{\Gamma(k+M)}{k ! \Gamma(M)}\left(\frac{\bar{k}}{\bar{k}+M}\right)^{k}\left(\frac{M}{\bar{k}+M}\right)^{M}
\label{eqn:Nb}
\end{equation}

Fitting this negative binomial distribution requires the extraction of photon counts from raw detector images, and works fairly well in the hard x-ray regime and for large pixel size detectors \cite{Hruszkewycz-2012-PRL,sikorski-jsr-2016,sun-2020-jsr}. In cases where the pixels are small, or the energy of the x-rays is much lower, this process can involve additional obstacles. One challenge is the point spread function of a single photon  can spread non-uniformly over many pixels. This is especially true in the soft x-ray regime, where there can be a large variability in the charge cloud size -- owing to variable diffusion lengths within a pixel -- and low signal to noise ratios. These effects have recently been shown to be corrected by a variational droplet model called the Gaussian Greed Guess (GGG) droplet model \cite{burdet2021absolute}, which can fit the large variation in charge cloud radii to produce discrete images where each pixel contains the number of corresponding photons. 

While 'droplet-type' models have been largely successful, there is a need to increase the speed of these computational models as well as to handle common scenarios, such as low signal-to-noise.
A few works have employed machine learning techniques to address some of these outstanding challenges. For instance, the use of convolutional neural networks to analyze XPCS data for well-resolved speckles has showed the denoising approaches are able to achieve significantly better signal-to-noise statistics as well as estimations of key parameters of interest \cite{campbell2021outlook,konstantinova2022machine,konstantinova2021noise}. Previous work has also considered the single-photon analysis for hard X-ray detectors using machine learning. One approach \cite{blaj2019ultrafast} has been to use a tensorflow computational graph with hand-crafted convolutional masks derived from an in-depth study of photon physics at semiconductor junctions \cite{blaj2017analytical}. This implementation is extremely fast, but does not apply to regimes where there may be a large number of photons per droplet. Another method \cite{abarbanel2019artificial} proposes a feed-forward neural network architecture, based on a sliding prediction of 5x5 regions of the input image. This was proposed for the photon map prediction task and is shown to be applicable for hard X-ray, low count rate experiments. However, additional factors such as noise, low photon energies, and insufficient signal-to-noise ratios can cause limitations in this methodology and thus obscure scientific results. Furthermore, in cases where a higher intensity can be measured, the charge clouds can quickly coalesce, making this problem intractable.

In this work, we expand the applicability of this ultrafast method  by demonstrating robust single-shot prediction using an AI-assisted algorithm in the soft X-ray regime for data with relatively high average count-rates and significant charge sharing. This is carried out using a fully convolutional neural network architecture \cite{long2015fully}, which we compare against the GGG method, currently the best algorithm for soft X-ray analysis using small pixel-size detectors \cite{burdet2021absolute}. 
We find that we are able to access a new phase space of measurement parameters that, until now, has not been accessible in structural dynamics studies using this method. Our algorithm enables a two order of magnitude speedup on appropriate hardware, is relative accurate for low contrast cases, and is stable at higher intensities than the GGG algorithm. We first describe the machine learning model and simulator used to train it, specifying the architecture, how the model is trained, and the evaluation metric used. This is followed by our main results, and the three areas which were shown to return excellent results relative to the current state-of-the-art models. Finally, we end with a discussion of uncertainty quantification, and how one can judge the error for different models.

\section{Modelling and Analysis Approach}

\subsection{Simulator Description}
One key issue in the development of supervised machine learning algorithms is a robust simulator which can adequately describe the data. To describe the simulator here,
we denote an input XPFS frame as $x_i \in \mathbb{R}^{90x90}$ and the corresponding output photon map as $p_i \in \mathbb{R}^{30x30}$. The 3x3 reduction in dimensionality between $x_i$ and $p_i$ is used to mimic the speckle oversampling factor that is typically used in LCLS experiments. The final calculations are performed on a 30x30 image to allow for the proper photon events to be expressed per speckle. 

\begin{figure}[h]
    \centering
    \subfigure[ ]{\includegraphics[width=0.4\textwidth]{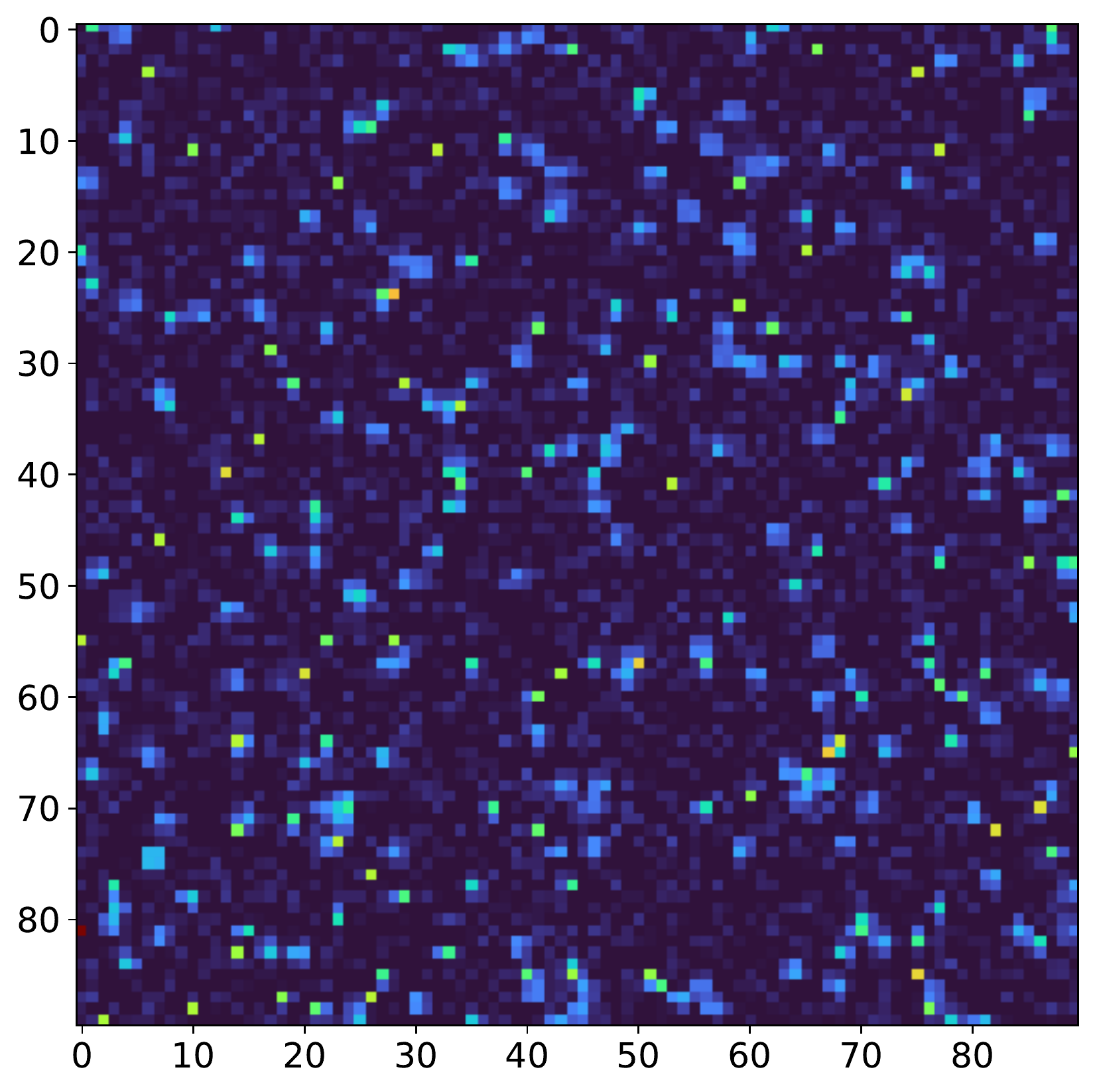}}
    \subfigure[ ]{\includegraphics[width=0.4\textwidth]{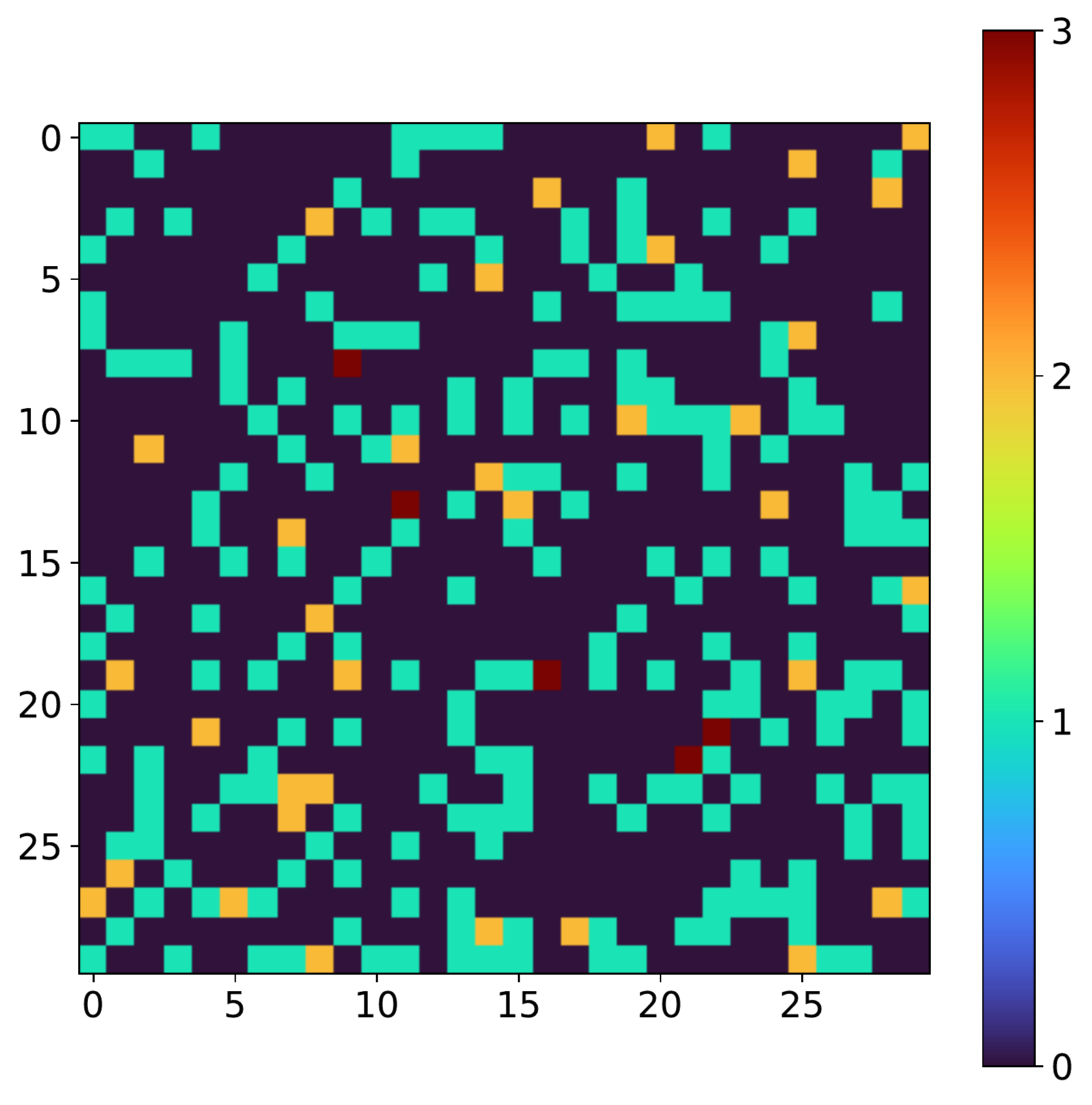}} 
    \caption{\label{fig:example_data} (a) An example of an XPFS detector image over a 90x90 pixelated detector. (b) Corresponding image of the photon map produced by the simulator for the detector image, plotted as the photon distribution per speckle.}
\end{figure}

The detector images and corresponding photon maps are simulated according to the exact parameters described in \cite{burdet2021absolute} which were tuned to mimic a previous experiment by matching the overall pixel and droplet histogram. To simulate ground truth photon maps, the following ranges were used: $\bar{k}$ $\in$ [0.025, 2.0] and $C\left(q,t\right)$ $\in$ [0.1, 1.0]. The relevant detector parameters are the probabilities ($w_i$) and sizes ($\sigma_G$) of the photon charge clouds, the variance of the zero-mean Gaussian background detector noise ($\sigma_N$) and the total number of analog-to-digital units (ADUs) per photon. These parameter values are reproduced below in Table~\ref{tab:simulation_params} and an example of a detector image / photon map pair is shown in Figure~\ref{fig:example_data}. For comparison, we used the Gaussian Greedy Guess (GGG) algorithm with relevant parameters which were optimized for these specific simulation parameters described above \cite{burdet2021absolute}.

\begin{table}[ht]
\centering
\begin{tabular}{|c|c|c|c|c|c|c|c|c|}
\hline
Number of pixels per droplet & 1 & 2 & 3 & 4 & 5 & 6 & 7 & 8 \\
\hline
$\sigma_{G}$  & 0.10 & 0.25 & 0.35 & 0.45 & 0.55 & 0.60 & 0.65 & 0.70 \\
\hline
$w_i$ & 0.20 & 0.125 & 0.125 & 0.175 & 0.175 & 0.10 & 0.05 & 0.025 \\
\hline
\multicolumn{4}{|c|}{photon ADU = 340} &  \multicolumn{5}{c|}{ $\sigma_N$ = [0-12] ADU} \\
\hline
\end{tabular}
\caption{\label{tab:simulation_params} Simulation parameters used to generate detector images based on previous XPFS data collected by Seaberg et al.~at the iron $L$-edge for magnetic scattering \cite{seaberg-2017-prl}.}
\end{table}

\subsection{Model Architecture}

\begin{figure}[h]
    \centering
    \fbox{\includegraphics[width=\textwidth, trim = 0cm 11cm 1cm 9cm, clip]{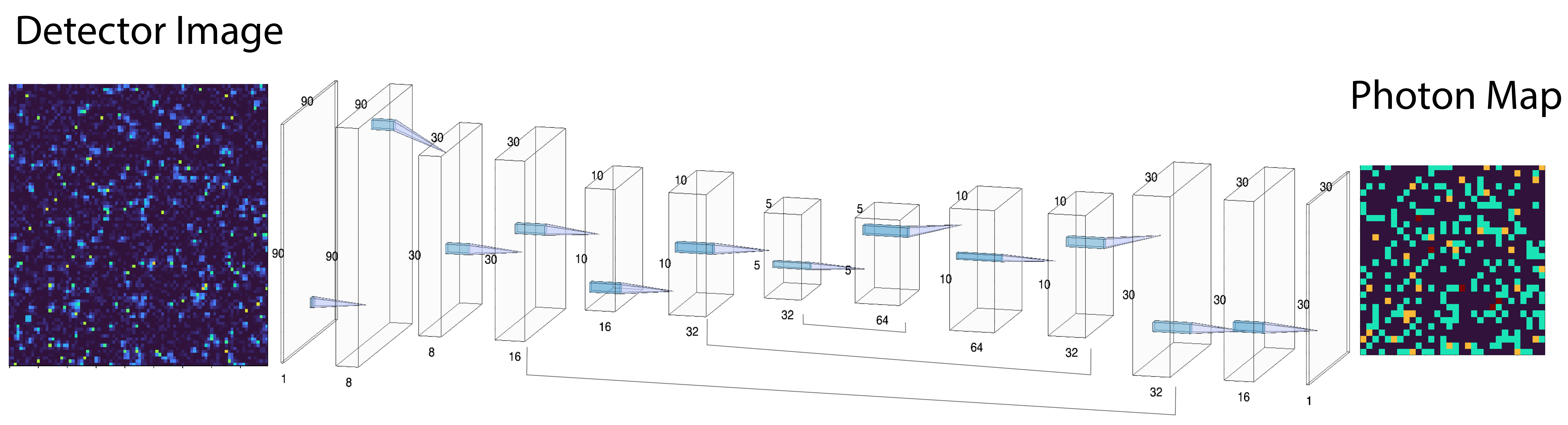}}
    \caption{A schematic for the U-Net neural network developed here for single photon counting detection. The input is given by the 90x90 detector image with output shown of the resultant 30x30 speckle photon map. Convolutional block layers are shown between the input and output images using NN-SVG \cite{lenail2019nn}. 
    \label{fig:model_architecture} }
\end{figure}

In this problem, we seek a supervised machine learning model which learns the functional mapping $f: x_i \mapsto p_i$, from $N_f$ paired simulated data points ($x_{i=1:N_f}, p_{i=1:N_f}$). In this case, the functional mapping was chosen to be a U-Net neural network (Figure~\ref{fig:model_architecture}), a fully convolutional autoencoder architecture, which is characterized by having skip-connections between different layers of resolution and has been shown to perform well on various image segmentation tasks\cite{ronneberger2015u}. In the schematic in Figure~\ref{fig:model_architecture}, the architecture is outlined via successive "convolutional blocks". Each such convolutional block consists of two convolutional layers sequentially applied to the input. After each convolution, we utilize batch normalization \cite{ioffe2015batch} to ensure robust optimization, followed by a Rectified Linear Unit (ReLU) activation.

\subsection{Training Details and Validation Metrics}

To train the model, we use the Frobenius norm between the predicted photon maps ($\hat{P}_{i}$) and the true photon maps ($P_{i}$). This loss function measures the average squared deviation between the predicted photon map  and the true photon map, where the average is taken both within a given frame (which contains $N_p$ pixels) and between frames ($N_f$) in the dataset. 
\begin{align}
L \left( P, \hat{P} \right) = \sum_{i=1}^{\text N_f} \sum_{j=1}^{\text N_p} \|P_{i,j} - \hat{P}_{i,j}\|^2_2
\end{align}

The U-Net model is trained by minimizing $L \left( P, \hat{P} \right)$ with respect to the model parameters. To train the neural network, we use the following hyperparameters: Adaptive Moment Estimation (ADAM) algorithm for optimization \cite{kingma2014adam}, batch size = 128, learning rate = 0.001, and batch normalization. We used NVIDIA A100 GPU hardware with the Keras API \cite{chollet2015keras}. 

We performed analysis at low and high count rates ($\bar{k}$) and trained respective models. For the low- $\bar{k}$ data, 100,000 training data points were simulated based on the detector parameters in Table~\ref{tab:simulation_params} and uniformly selecting $\bar{k}$ in range [0.025, 0.2] and $C\left(q,t\right)$ over the range [0.1, 1.0]. For the high- $\bar{k}$ analysis, 300,000 data points were used for training with an equal proportion of datapoints coming from $\bar{k}$ $\in$ [0.025, 0.2], [1.0, 2.0] and [0.025, 2.0], respectively, with the the $C\left(q,t\right)$ randomly chosen from the range [0.1, 1.0].  
To select between competing trained models, the optimal neural network was selected based on maximizing the correlation between the estimated and the true contrast on held-out validation sets of size 5000 for contrast values in the range [0.1, 1.0] with increments of $0.05$. Here, it is worth emphasizing that the metric used to evaluate the photonizing task is important. For example, the overall accuracy is not necessarily a good metric since many photon maps have a small number of photons. Therefore, a model which uniformly predicts $0$ for each pixel will show a uninformatively high accuracy, which is clearly not the desired performance and will lead to poor statistics. Similar issues have been documented in problems with high class imbalances \cite{luque2019impact}, and correlation based similarity metrics for evaluation are recommended therein \cite{herlocker2004evaluating}. Since our final goal is to obtain a good estimate of the contrast, it is useful to use this information directly in the evaluation metric. For the low- $\bar{k}$ analysis, validation datasets were simulated in the range [0.025-0.2]. For the high- $\bar{k}$ analysis, two additional datasets corresponding to $\bar{k}$ in the ranges [1.0, 2.0] and [0.025, 2.0] were used. The evaluation metric for this analysis was the average correlation for the three different $\bar{k}$ ranges. An example of a sample validation plot is shown in Appendix A.

Finally, to obtain an estimate of the contrast from the predicted photon maps we use the maximum likelihood estimation procedure on the negative binomial distribution. In general, the negative binomial distribution is a function of both $C\left(q,t\right)$ and $\bar{k}$. However, we directly use a per-image estimate for $\bar{k}$ and therefore the MLE procedure reduces to a 1D optimization in $C\left(q,t\right)$ \cite{roseker2018towards}. 

\section{\label{sec:resuls}Results and Analysis}

\subsection{Speed of Inference}

As X-ray sources and detectors move towards faster repetition rates nearing 1\,MHz, it is important to preserve the possibility of live data analysis. Here, we compare the speed of an optimized GGG droplet algorithm \cite{Burdet2021} against the trained CNN model (see Table~\ref{tab:SpeedML}). On 1 CPU, the CNN outperforms the GGG algorithm by roughly an order of magnitude. This advantage stretches to two orders of magnitude when comparing GGG parallelized across multiple CPU cores to the CNN running on one NVIDIA A100 GPU.  

\begin{table}[h]
\begin{center}
\caption{\label{tab:SpeedML} Speed Comparison Between the CNN and the GGG Algorithm. Rates are reported for a prediction on 1000 XPFS shots. Using a CNN deployed on GPU hardware yields a speedup of two orders of magnitude relative to a multi-CPU Droplet Implementation.}
\begin{tabular}{cccc}
Device & Algorithm & Rate (kHz) & Rate relative to 1 CPU \\
\hline
1 CPU & GGG & 0.008 & 1 \\
12 CPU & GGG  & 0.05 & 6\\
32 CPU & GGG  & 0.1 & 12 \\
1 CPU & CNN & 0.2 & 27 \\
1 GPU & CNN & 5.0 & 700 \\
\end{tabular}
\end{center}
\end{table}

\begin{figure}[h]
    \centering
    \includegraphics[width=0.6\textwidth]{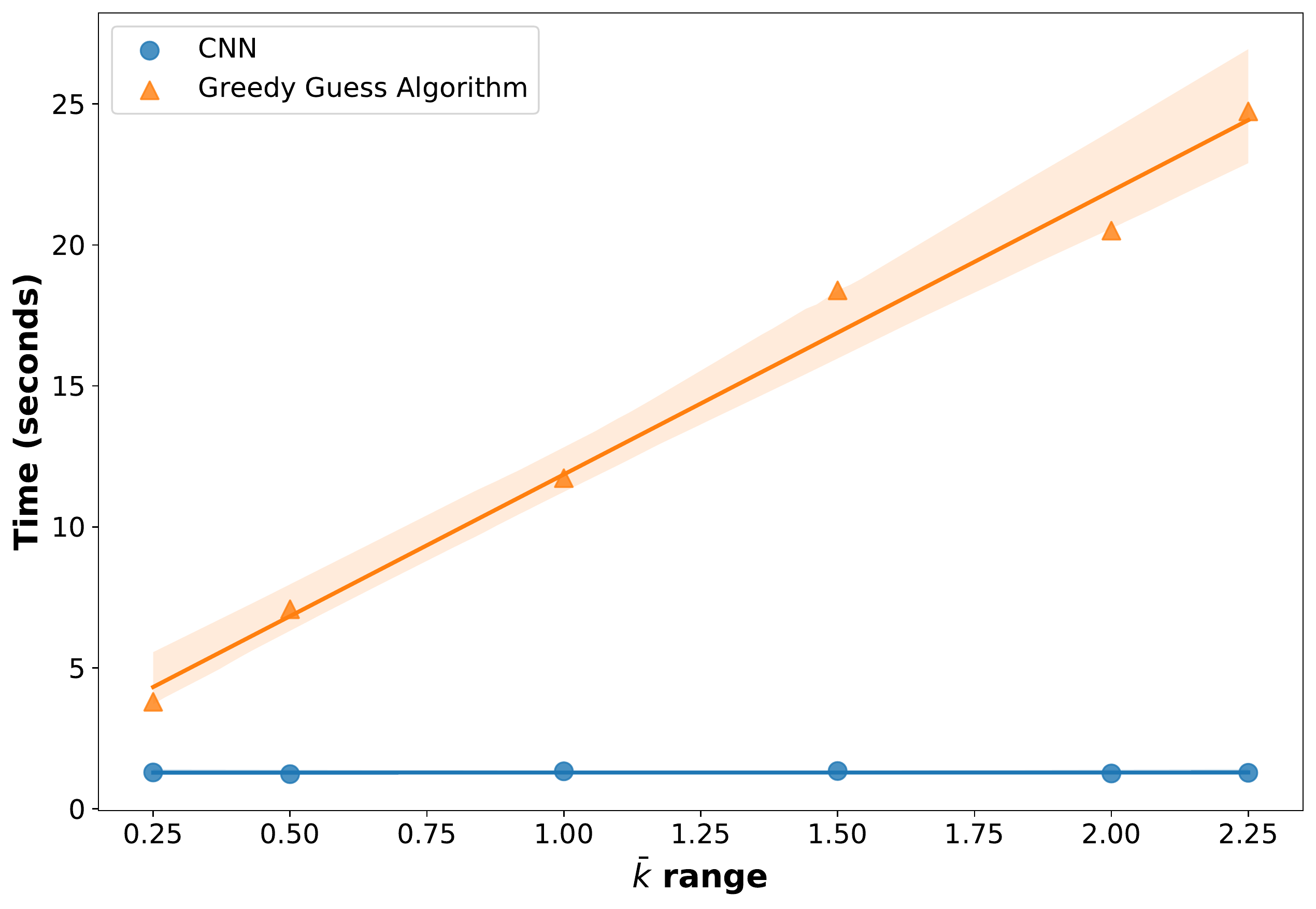}
    \caption{\label{fig:scaling} Time to make predictions on 2000 XPFS frames for the CNN vs. the GGG algorithm as a function of $\bar{k}$. Error bars are obtained from the standard error in the slope of the linear regression fit. The CNN exhibits constant scaling with $\bar{k}$ while the GGG scaling is observed to be linear.}
\end{figure}

The observed speedup presented here is consistent with intuition. At inference, the trained neural network, which consists primarily of matrix multiplication operations, is efficiently parallelized over thousands of GPU processes \cite{kirk2016programming}. In contrast, the GGG algorithm requires for-loop operations at the level of each droplet. For this reason, one additional beneficial property of the CNN model is that the prediction rate does not depend on the content of the XPFS frames and is consequently independent of $\bar{k}$. In contrast, for the GGG algorithm, the run time scales linearly with $\bar{k}$ (Figure~\ref{fig:scaling}). Here, it is worth mentioning that the GGG algorithm is already orders of magnitudes faster than the Droplet Least Squares algorithm \cite{Hruszkewycz-2012-PRL}, which is exponential in computational complexity.

\begin{figure}[h]
\begin{center}
    \subfigure[ ]{\includegraphics[width=0.46\textwidth]{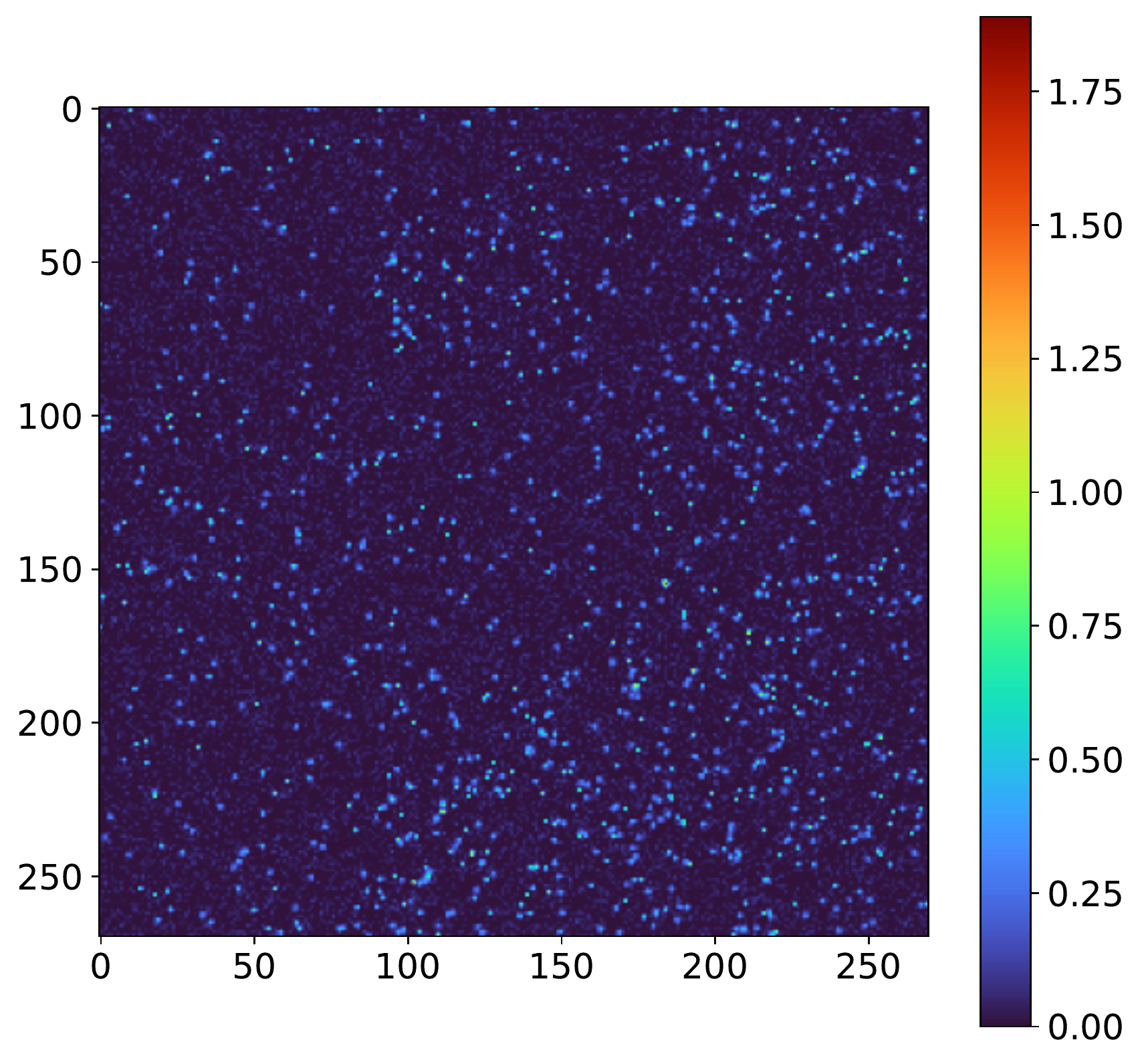}} 
    \subfigure[ ]{\includegraphics[width=0.43\textwidth]{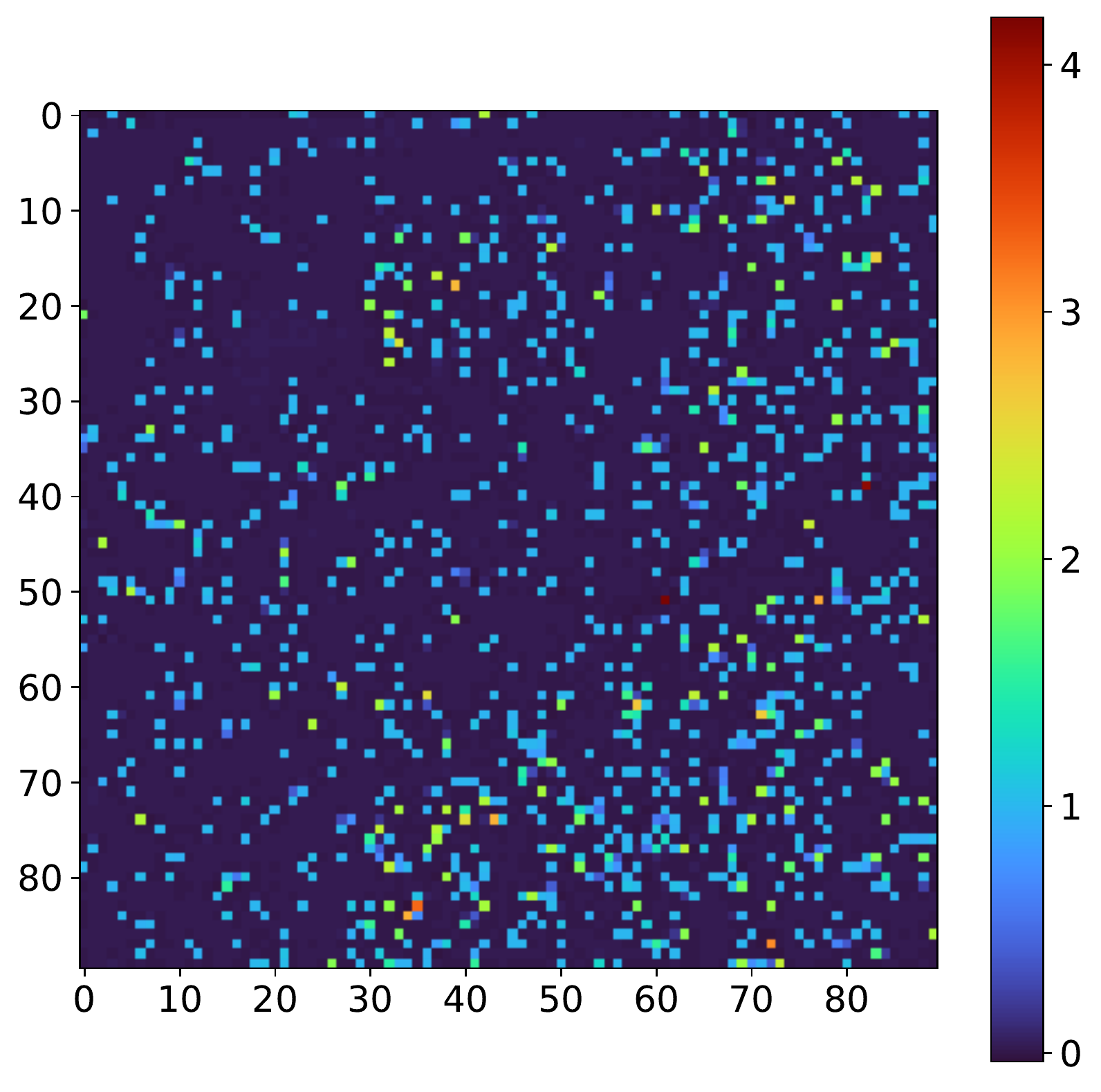}} 

\caption{\label{fig:size_scaling} (a) A larger input detector image with 270x270 pixels and (b) corresponding predicted photon map (90x90 pixels). The CNN is able to make predictions on larger inputs than it was trained on.}
\end{center}
\end{figure}

Finally, since the neural network architecture follows a fully convolutional paradigm, it is possible to make predictions on large input / detector sizes than those used in the training set. This is enabled by the fact that fully-convolutional architectures learn local spatial filters which apply to the full image, making the learning process relatively efficient. For this analysis, the neural network can handle input sizes of ($N_f$, 90$a$, 90$b$, 1), where $a$, $b$ are positive integers and $N_f$ denotes a variable number of frames. Note, that this allows for any detector size to be used after zero-padding to nearest (90$a$, 90$b$) frame size. We calculate the average time to make predictions on datasets of dimensionality [100, 90, 90, 1], [100, 270, 270, 1] and [100, 900, 900, 1]. We observe rates of 3.4, 3.1 and 0.3 kHz, respectively. As the rate only decreases a factor of 10 between a frame size of (90,90) and (900,900), it appears that we do not observe quadratic scaling that would be observed using droplet-based algorithms. Furthermore, the ability to analyze such data in a single-shot manner is a significant advantage over the sliding approach for droplet analysis which has been developed \cite{abarbanel2019artificial}. A representative example of the CNN prediction using an input resolution of 270x270 pixels is shown in Figure~\ref{fig:size_scaling}. 

\subsection{Accuracy}

In this section, we compare the prediction quality of the CNN model against the GGG algorithm on data which simulates LCLS experiments and for which the ground truth contrast is known \cite{burdet2021absolute}. We begin with an analysis of low count-rate ($\bar{k}$ $\in$ [0.025, 0.2]) data and subsequently present results for higher count-rate data ($\bar{k}$ $\in$ [0.2, 2.0]). To quantify performance, we show parity plots for the predicted and true contrast on datasets with varying contrasts. 

\begin{figure}[h]
    \centering
    \subfigure[ ]{\includegraphics[width=0.43\textwidth]{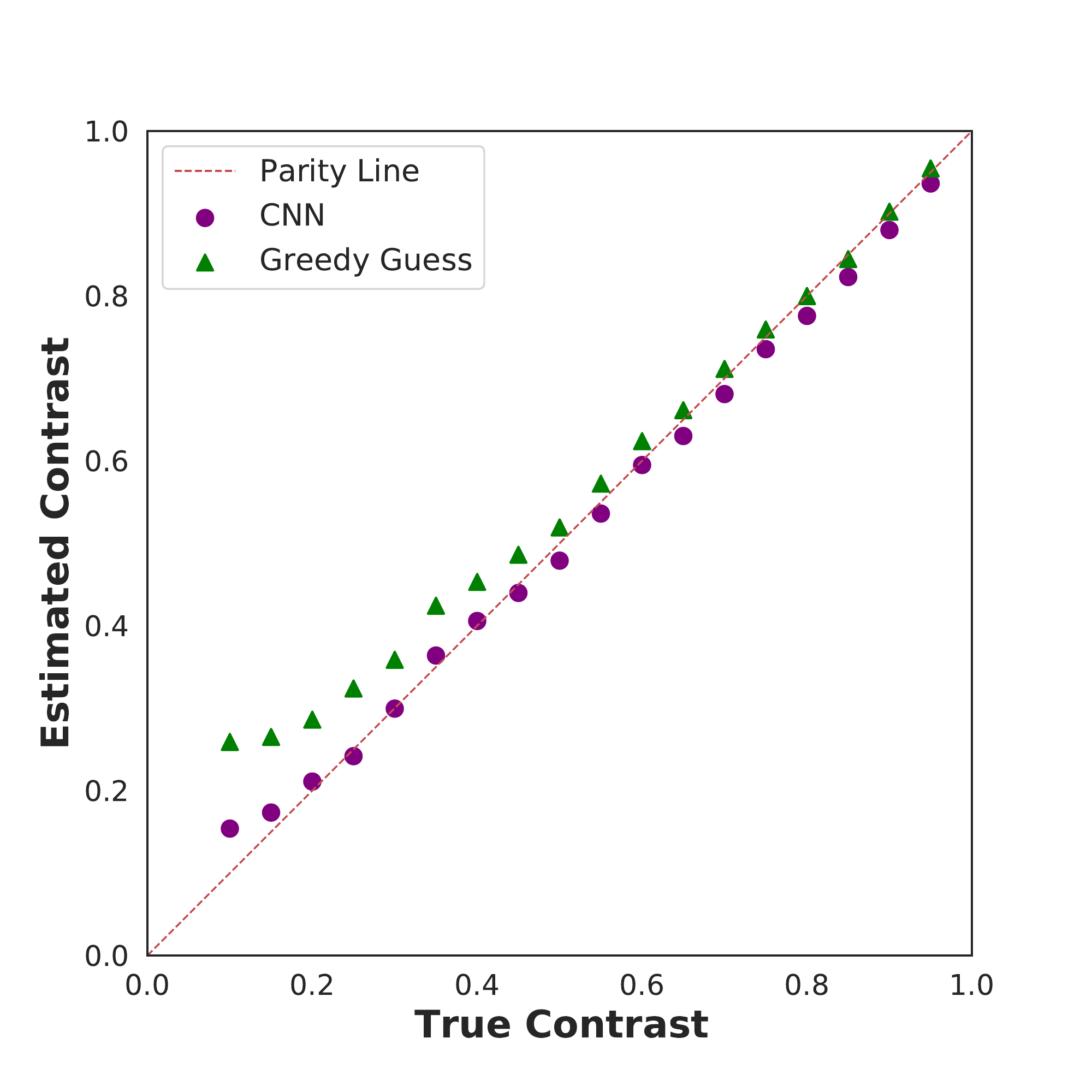}}
    \subfigure[ ]{\includegraphics[width=0.43\textwidth]{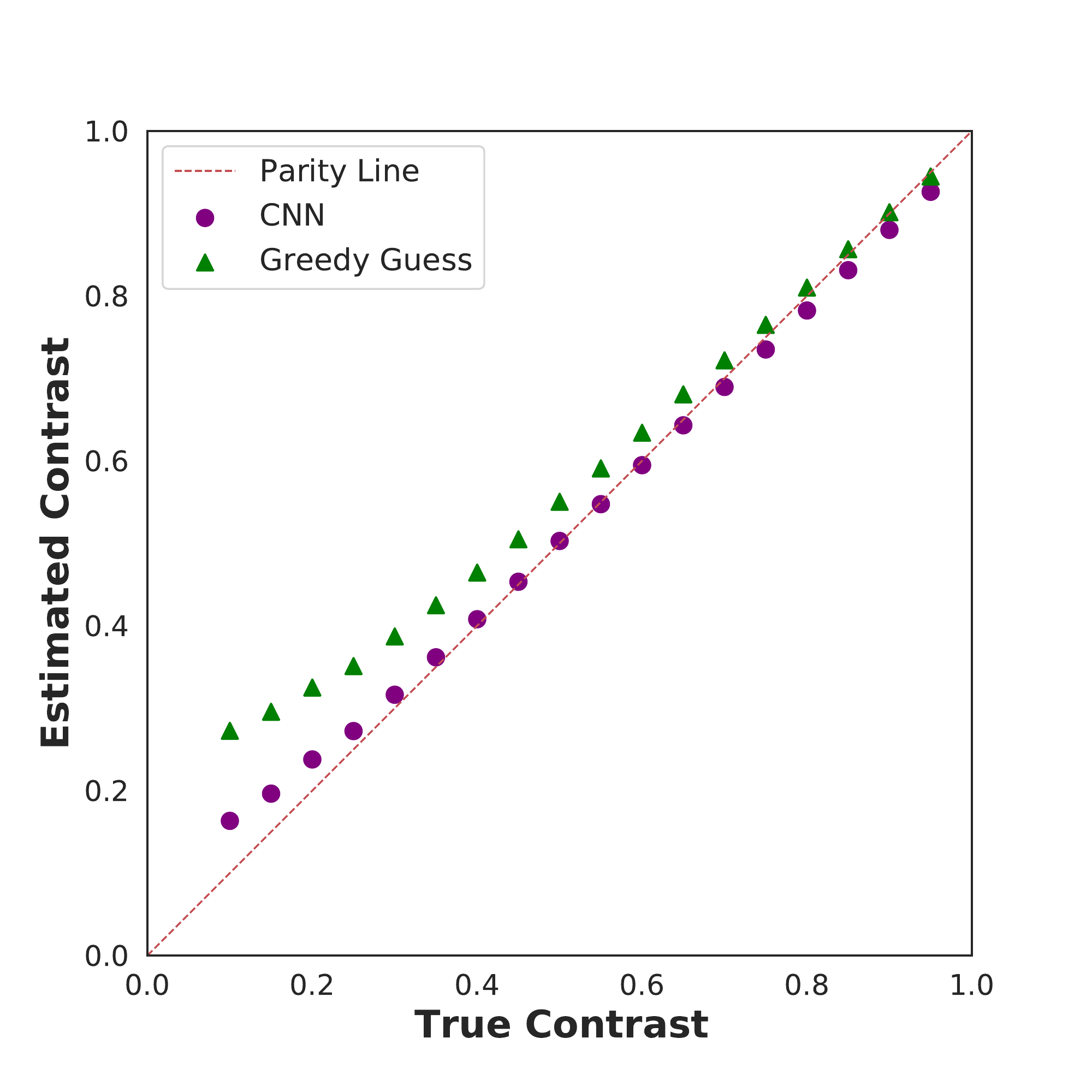}} 
    \subfigure[ ]{\includegraphics[width=0.43\textwidth]{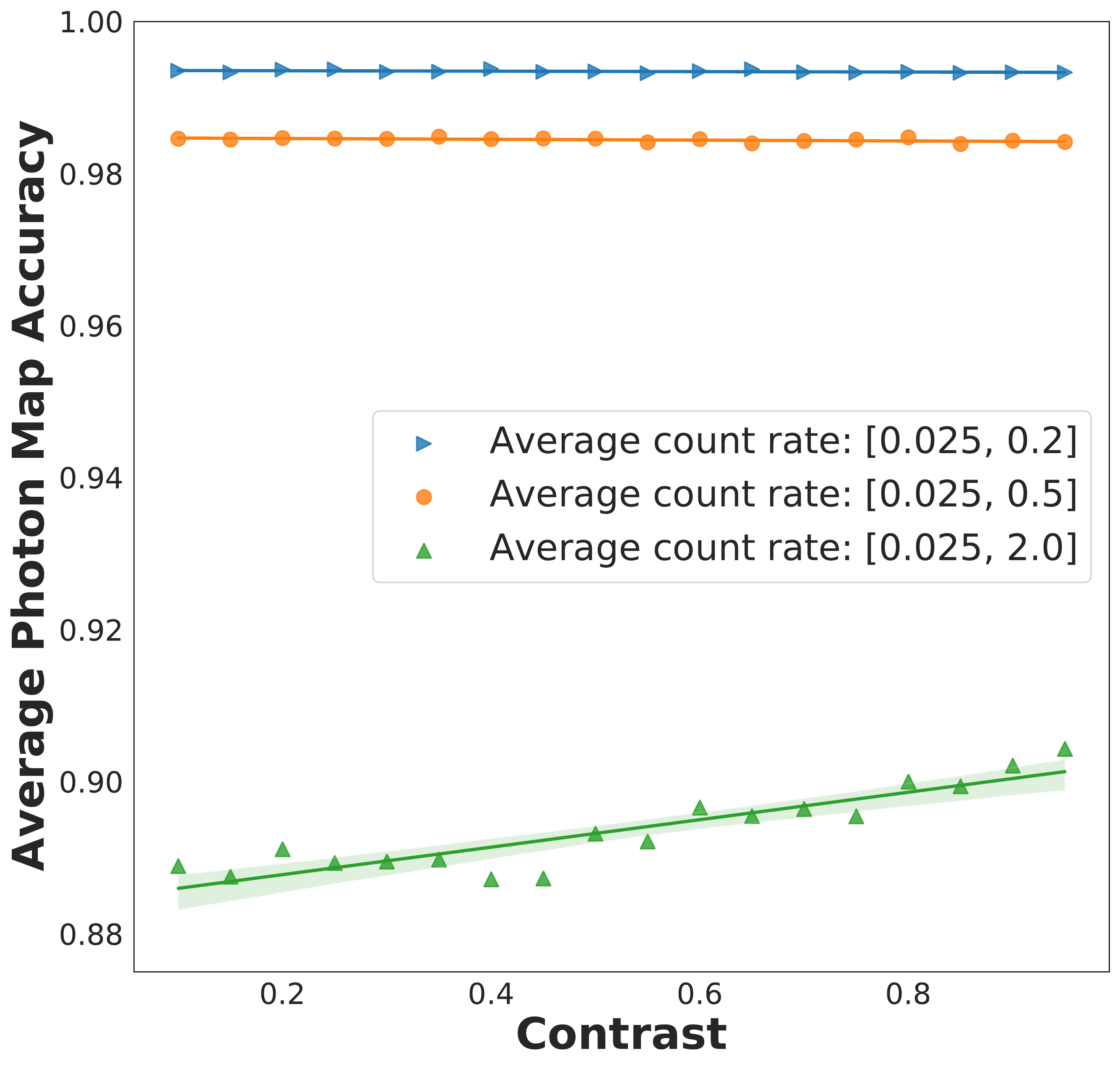}} 
    \subfigure[ ]{\includegraphics[width=0.46\textwidth]{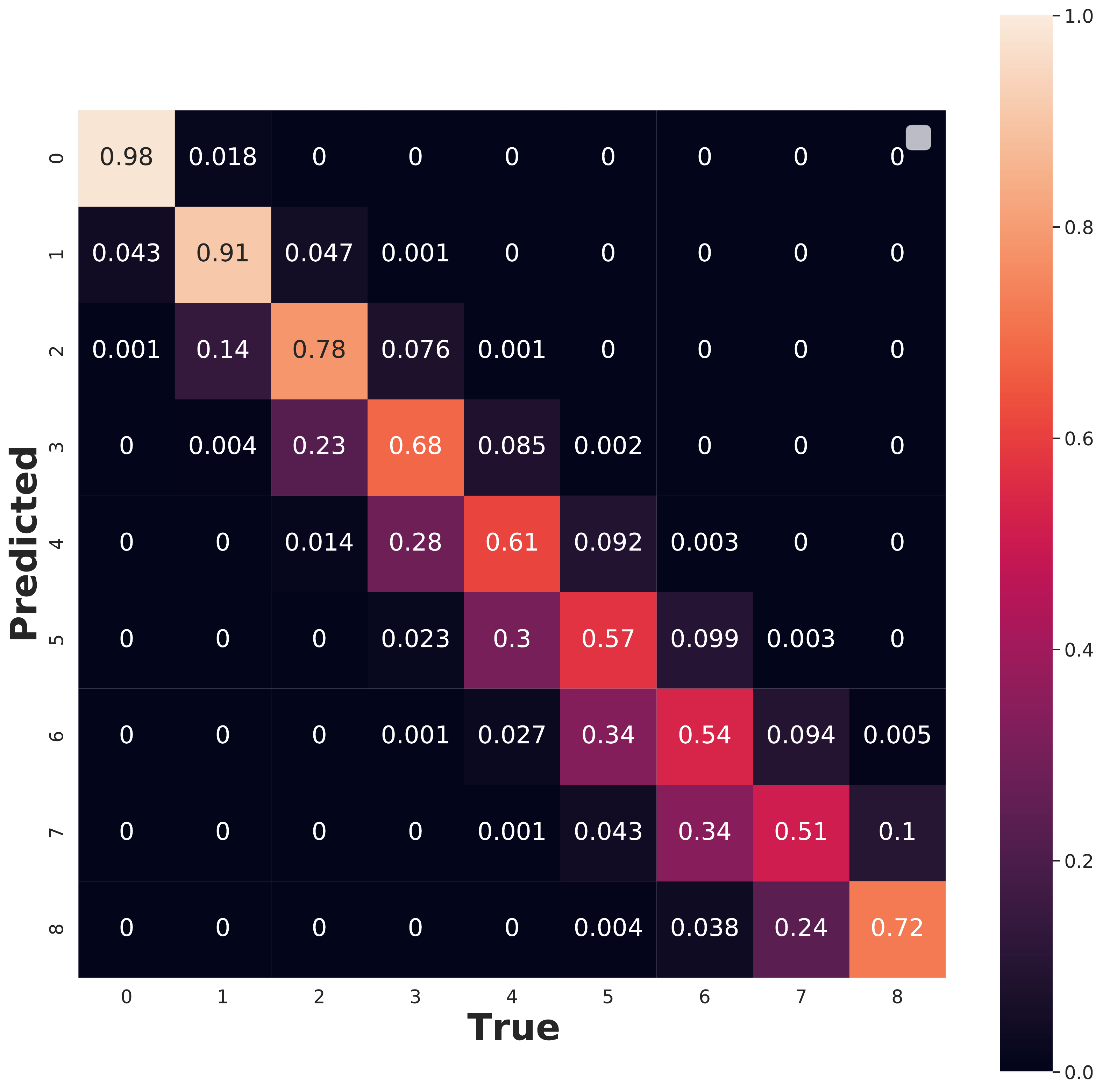}} 
    \caption{\label{fig:errors} Predicted Contrast versus the True Contrast for CNN and GGG algorithms for (a) $\bar{k}$ $\in$ [0.025, 0.2] and (b) $\bar{k}$ $\in$ [0.025, 2.0]. Each datapoint corresponds to prediction on a dataset of 2000 datapoints and subsequent maximum likelihood estimation for the contrast parameter. Notably, the CNN model exhibits much smaller bias than the GGG algorithm at low contrasts and high $\bar{k}$. (c) Average accuracy as a function of contrast level for three separate $\bar{k}$ ranges. Error bars are obtained from the standard error in the slope of the linear regression fit. (d) Confusion matrices for CNN predictions on testing data with $\bar{k}$ $\in$ [0.025-2.0]. The CNN model makes a higher proportion of errors on high photon count events and asymmetrically under-predicts high photon events.}
\end{figure}

At low $\bar{k}$ and high contrasts, the CNN and GGG algorithm give good predictions for the contrast. Although, it is worth pointing out that in this regime the CNN algorithm systematically underpredicts the contrast and has a slightly larger bias than the GGG algorithm. However, at low contrast levels, the GGG algorithm exhibits much greater bias than the CNN (Figure~\ref{fig:errors}a). One possible reason for the overall superior performance of the CNN is that it takes into account variation in photon charge cloud sizes during training. Here, it is worth emphasizing that even after optimizing droplet parameters on simulated data with known detector parameters, the GGG algorithm still exhibits large bias for lower contrast values, indicating that the algorithm may not have the complexity required to fully treat such data. 

As $\bar{k}$ increases, the CNN performance decreases on an average photon map accuracy basis (Figure~\ref{fig:errors}c). To further examine the CNN errors, we clipped the output photon map to the range of [0,8] (i.e. no photon map has more than eight photons or less than zero photons) and analyzed the confusion matrix (CF$_{i,j}$) of the predictions (Figure~\ref{fig:errors}d). The diagonal of the confusion matrix represents per-class accuracy. For instance, CF$_{2,2}$ represents the accuracy of prediction for pixels containing two photons. By examining the diagonal elements, it is clear that the CNN model makes a greater proportion of errors for higher photon counts; note the trend does not hold for the eight photon event due to the clipping operation. The off-diagonals of the confusion matrix indicate how the model makes errors. For example, the CF$_{3,6}$ term indicates the probability of the model assigning three photons to a pixel when the true number of photons was actually equal to six. From these elements, we see that the CNN tends to systematically under-predict high photon events. Taken together, these observations suggest that there is a dataset imbalance issue owing to the fact that low photon events are more probable in the training set. 

Although at high $\bar{k}$, the CNN is less accurate on a per-photon map basis (relative to its low $\bar{k}$ performance), this does not necessarily imply inferior contrast predictions. In fact, the parity plots are similar for low and high $\bar{k}$ cases (Figure~\ref{fig:errors}b). This observation stems from the trade-off between information content and accuracy at high $\bar{k}$ (Appendix B). For a fixed dataset size, it is harder to estimate photon counts correctly, but the counts have significantly more information about the unknown $C\left(q,t\right)$ parameter. 

\clearpage 

\begin{figure}[h]
    \includegraphics[width=0.6\textwidth]{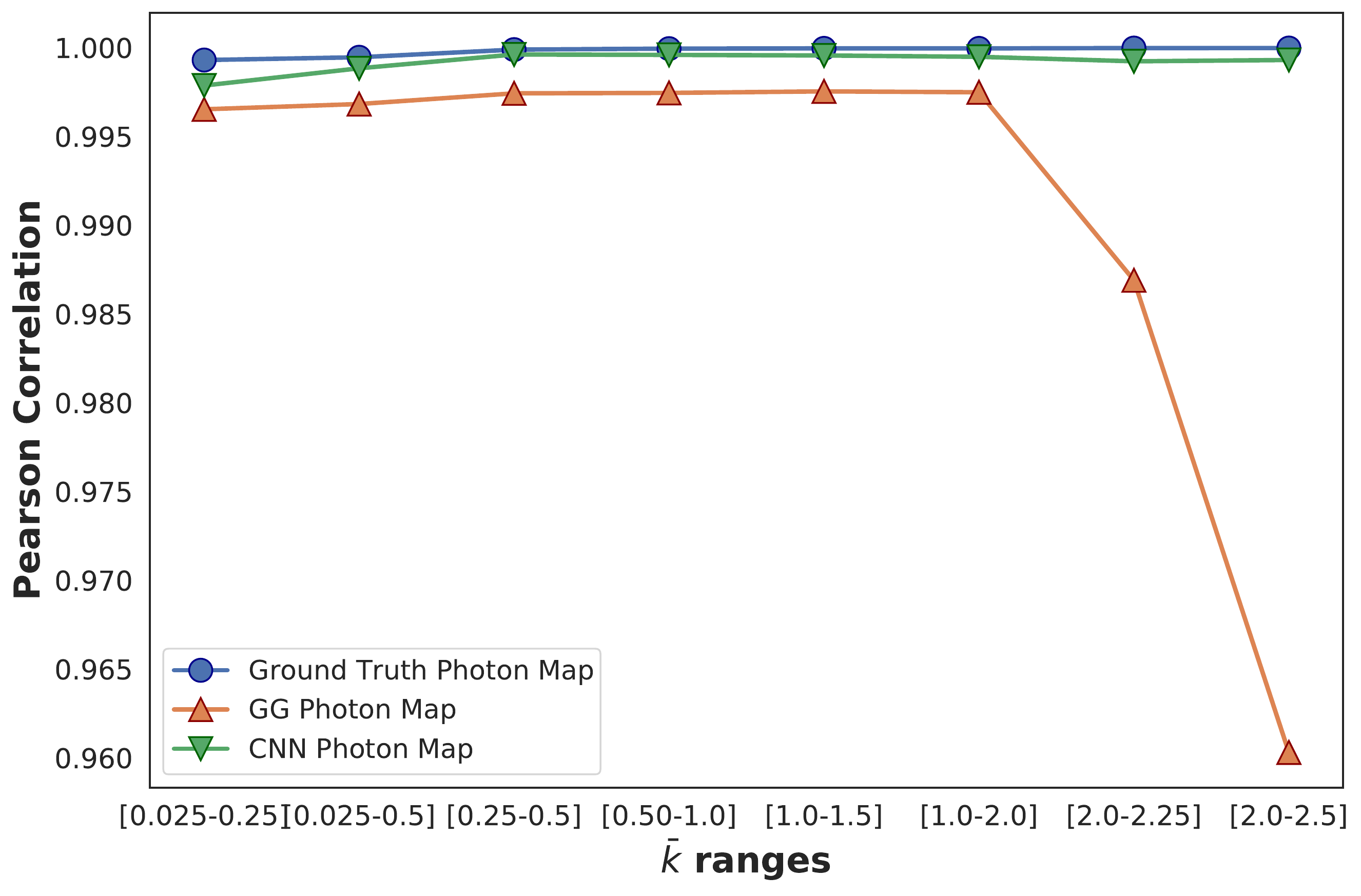} 
    \caption{\label{fig:correlation}Contrast-contrast parity correlation for datasets generated using different $\bar{k}$ ranges. Note, the ground truth photon maps do not have perfect correlation due to finite sampling statistics.}
\end{figure}

In Figure~\ref{fig:correlation}, we quantify the performance of the CNN model and the GGG algorithm at different $\bar{k}$ ranges using the correlation in the contrast-contrast parity plot as our metric. It is evident that the GGG algorithm is slightly biased across all $\bar{k}$ levels and performs poorly for $\bar{k} > 2.0$. This is unsurprising, as droplet-based algorithms were designed to cope with small droplets with relatively few overlapping charge clouds. Furthermore, this implies further development of the CNN algorithm will be capable of handling large $\bar{k}$ data sets.

\subsection{Uncertainty Quantification}
In this section, we consider a neural network ensemble approach to quantify the uncertainty in the predicted photon maps and contrasts. The motivation for such an analysis is that, while deep learning models have exhibited significant successes in their application to scientific problems, they have a tendency to engender overconfident predictions that may be inexact. As an example, neural networks are unable to recognize Out Of Distribution (OOD) instances and habitually make erroneous predictions for such cases with high confidence \cite{amodei2016concrete,nguyen2015deep, hendrycks2016baseline}. In reliability-critical tasks, such errors and uncertainties in model predictions have led to undesirable outcomes \cite{NTSB1, NTSB2, NTSB3}. In this context, quantifying the uncertainties in deep learning model predictions is highly desirable. 

\begin{figure}[h!]
\begin{center}
\includegraphics[width=0.6\textwidth]{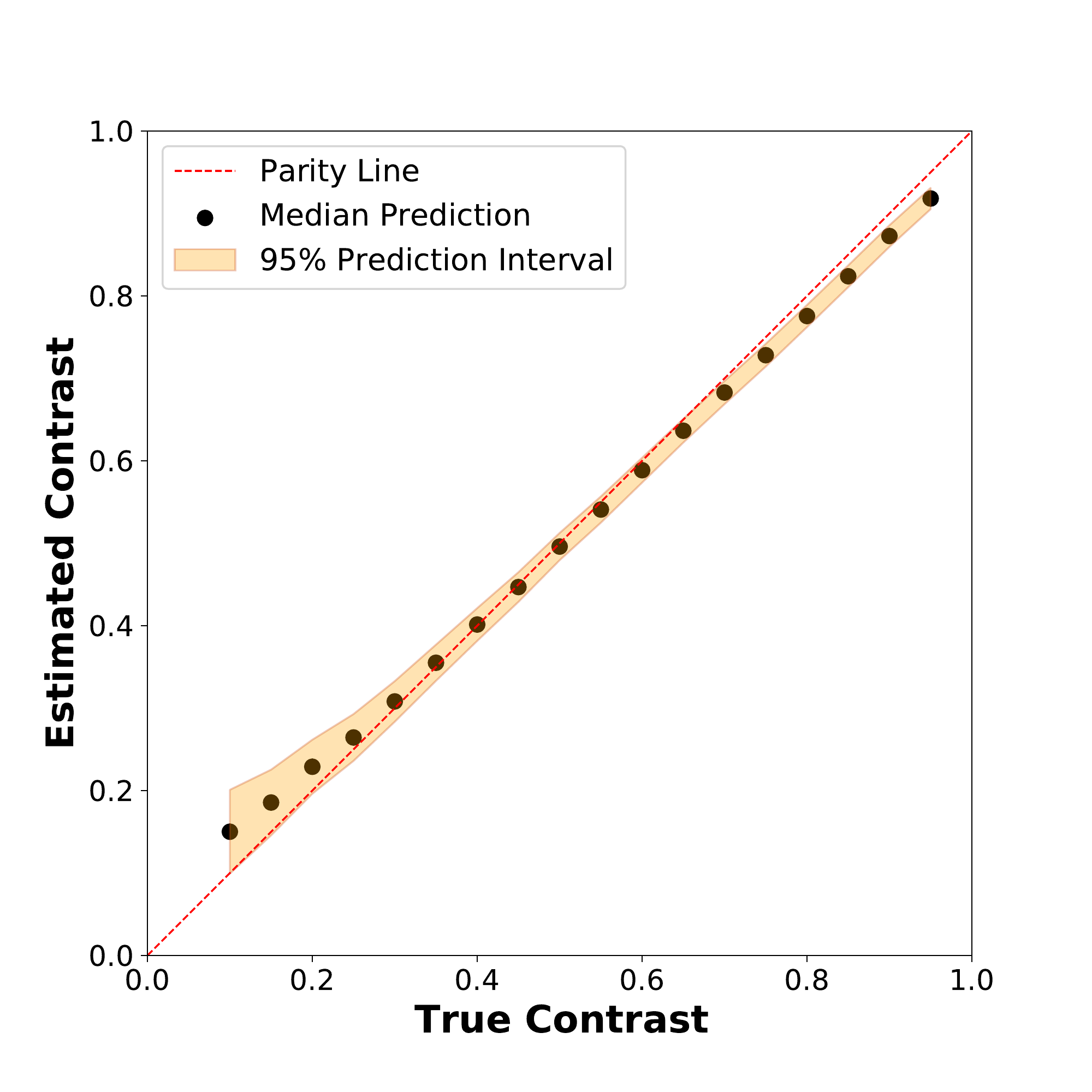}
\caption{\label{fig:t_error_bars} a) Contrast-contrast parity plot for data in the $\bar{k}$ range of [0.025, 2.0] using a vertical neural network ensemble. The contrast point prediction is obtained from the median contrast prediction and the $95\%$ prediction intervals follows from assumed $t$-distribution statistics.}
\end{center}
\end{figure}

There are two sources of predictive uncertainty that need to be considered: Epistemic and Aleatoric. Epistemic uncertainty \cite{smith2013uncertainty} (reducible or subjective uncertainty) arises due to lack of knowledge regarding the dynamics of the system under consideration, or an inability to express the underlying dynamics accurately using models. Epistemic uncertainties can lead to biases in the predictions. Aleatoric uncertainty \cite{smith2013uncertainty} (irreducible uncertainty or stochastic uncertainty) arises due to noise in the training data, projection of data onto a lower space, absence of important features, etc. Aleatoric sources can lead to variances in the predictions.

For our analysis, we use an ensemble of neural networks to make a point prediction of the contrast $C\left(q,t\right)$ as well as to give an estimate of statistical uncertainty. This is in line with model ensembling based uncertainty quantification (UQ) methods validated in literature \cite{heskes1996practical,efron1992bootstrap}. Such ensembling accounts for aleatoric uncertainties due to the data and weight uncertainties. In our investigation, the neural network ensemble is formed via sequential sampling, wherein ten partially decorrelated models were sampled during the model training. Contiguous samples were spaced by ten optimization epochs each. The contrast is calculated for each model via a maximum likelihood procedure and the contrast point prediction is taken as the median predicted value. To estimate the model uncertainty, we provide a $95\%$ contrast prediction interval using the standard deviation of the predicted contrasts and making the assuming that the predictive distribution follows a $t$-distribution with nine degrees of freedom (Figure~\ref{fig:t_error_bars}). 

We see that the error bars are larger at lower contrasts and correctly captures the notion that the prediction task is harder at lower contrasts \cite{Burdet2021}. We also notice a systematic bias in the CNN models at high contrasts. This bias may arise due to the epistemic uncertainties due to the model form (structural uncertainty). Such structural uncertainties in deep learning models cannot be accounted for by any extant approach, including the procedure used in this investigation. However, this indicates a need for more refinement on the present approach (for instance, more fine grained optimization of the model architecture, etc) and will be explored more fully in future work.  

\begin{figure}[h]
\begin{center}
    \subfigure[ ]{\includegraphics[width=0.4\textwidth]{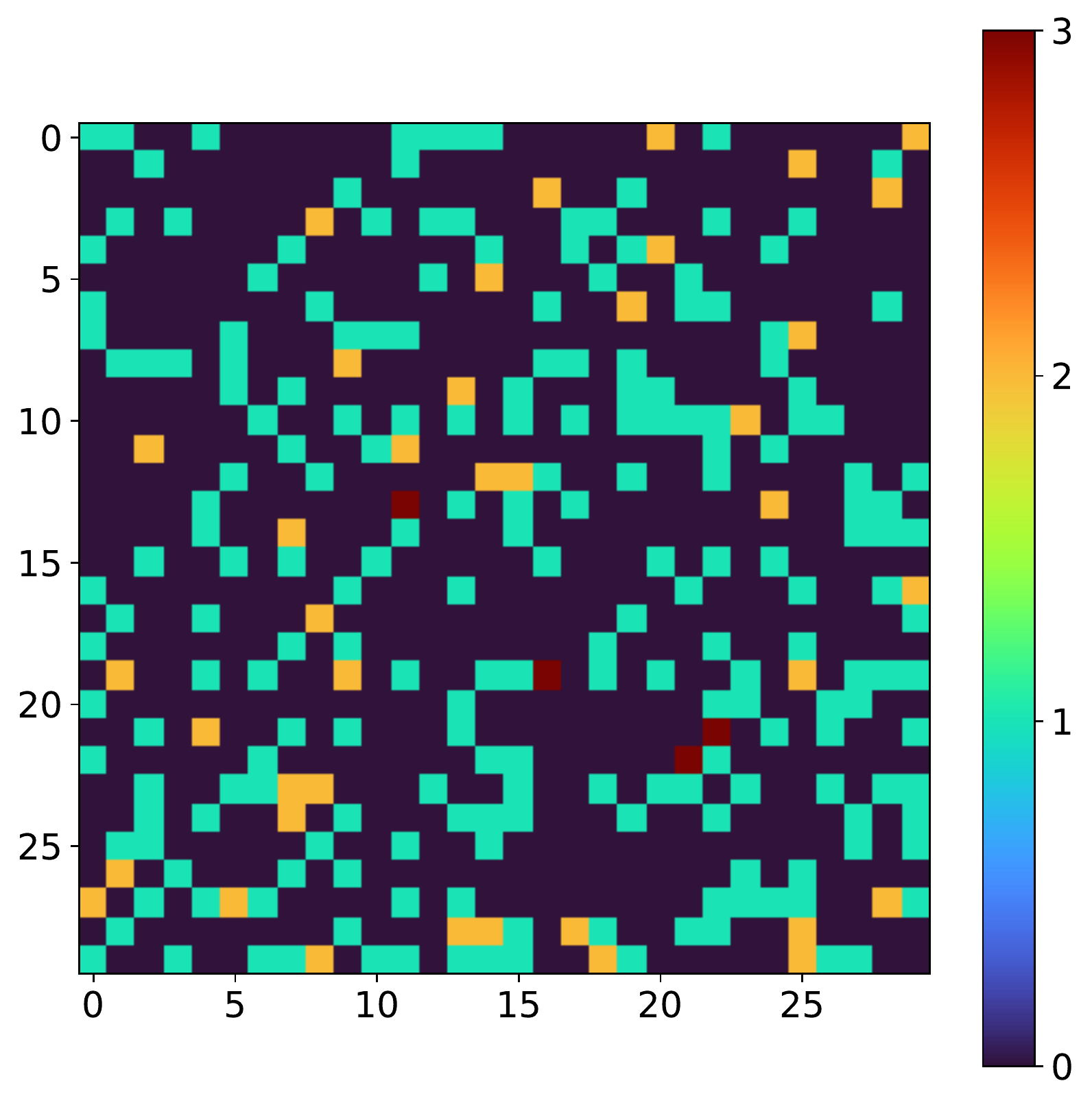}}
    \subfigure[ ]{\includegraphics[width=0.4\textwidth]{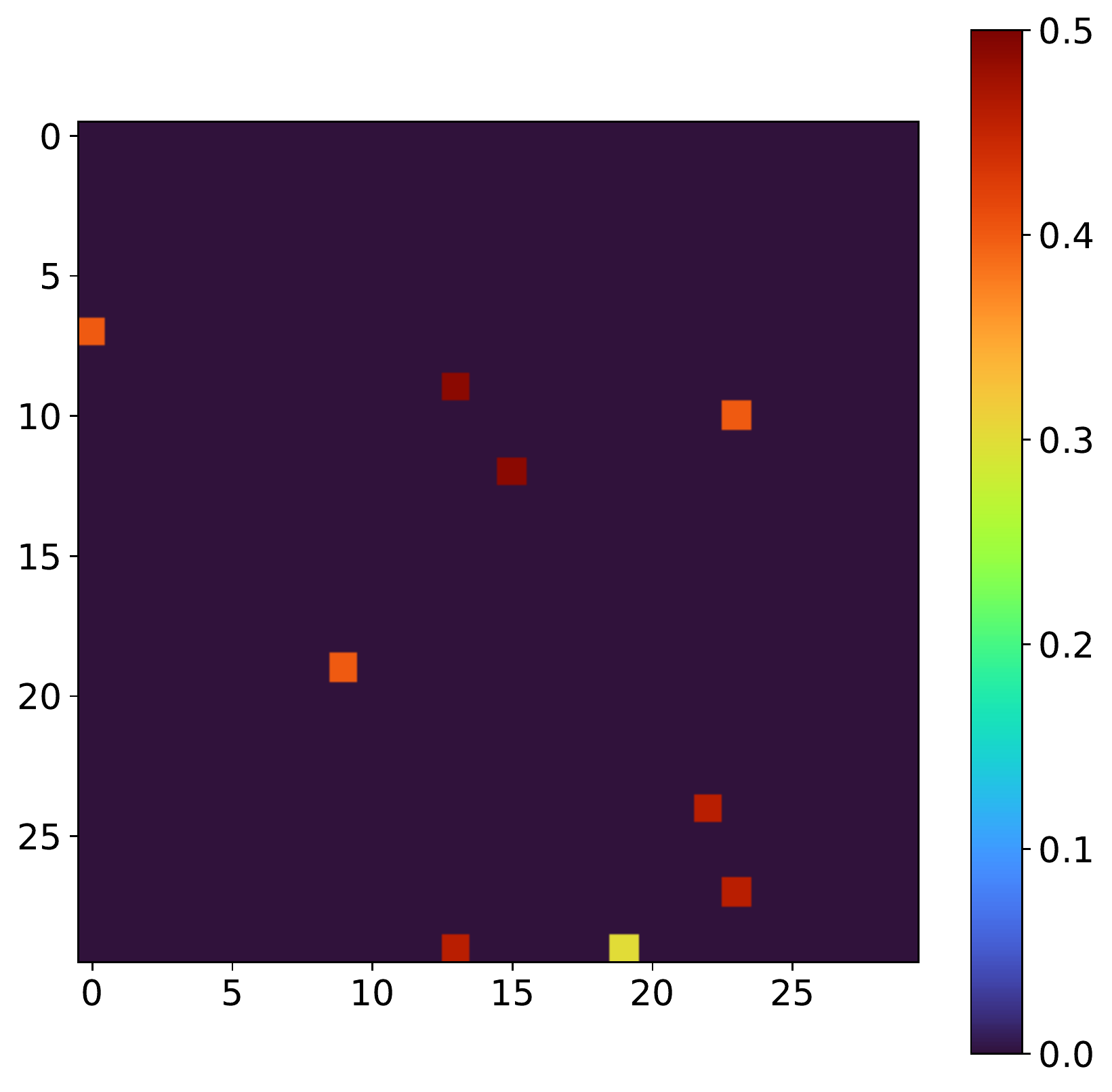}}
    
\caption{a) CNN ensemble used to visualize the (a) median prediction photon map and (b) the standard deviation prediction in the photon map. This analysis allows users to visualize the region of the image where the photon assignment may be challenging.}
\label{fig:mean_std}
\end{center}
\end{figure}

Another interesting avenue is to look at the median predicted photon map and the predicted standard deviation map to examine where the neural networks lack consensus. An example of this pair of outputs are shown in Figure~\ref{fig:mean_std}. Evidently, the ensemble predicts insignificant uncertainty for the majority of the image with the exception of a few pixels with relatively high uncertainty. An interesting future strategy could involve using the CNN model as a fast, initial approach and subsequently run more complex fitting algorithms on regions of the image with high predicted uncertainty.  

\section{Conclusions}

In this work, we have developed a convolutional neural network architecture which is capable of analysing single-photon X-ray speckle data in non-optimal situations, such as for small pixel size detector or with soft x-ray energies. We have benchmarked this algorithm on realistic simulated data and found that it outperforms the conventional Gaussian Greedy Guess (GGG) droplet algorithm in terms of speed and computational complexity. Furthermore, the algorithm is able to extract the contrast information for new ranges that were previously inaccessible, such as low contrast -- relevant for systems which scatter weakly, as well as in a high $\bar{k}$ regime. Both of the latter developments will create new opportunities to study fluctuations using XPFS in novel systems, such as in quantum or topological materials. 

\begin{acknowledgments}
This work is supported by the U.S. Department of Energy, Office of Science, Basic Energy Sciences under Award No. DE-SC0022216, as well as under Contract DE-AC02-76SF00515 for the Materials Sciences and Engineering Division. The use of the Linac Coherent Light Source (LCLS), SLAC National Accelerator Laboratory, is also supported by the DOE, Office of Science under contract DE-AC02-76SF00515. This work was also supported in part by funding from Zoox, Inc. J. J. Turner acknowledges support from the U.S. DOE, Office of Science, Basic Energy Sciences through the Early Career Research Program.
\end{acknowledgments}

\section*{Permissions}

The following article has been submitted to Structural Dynamics. 

\section*{Data Availability}

The data\cite{chitturixpfs} that support the findings of this study are openly available at \\ \mbox{\url{https://doi.org/10.5281/zenodo.6643622}}. Machine learning models are available at \url{https://github.com/src47/CNN_XPFS} upon manuscript acceptance. 

\section*{Conflict of Interest}

The authors have no conflicts to disclose.

\clearpage

\appendix

\section{Model Validation Metric}

\begin{figure}[h]
    \centering
    \includegraphics[width=0.6\textwidth]{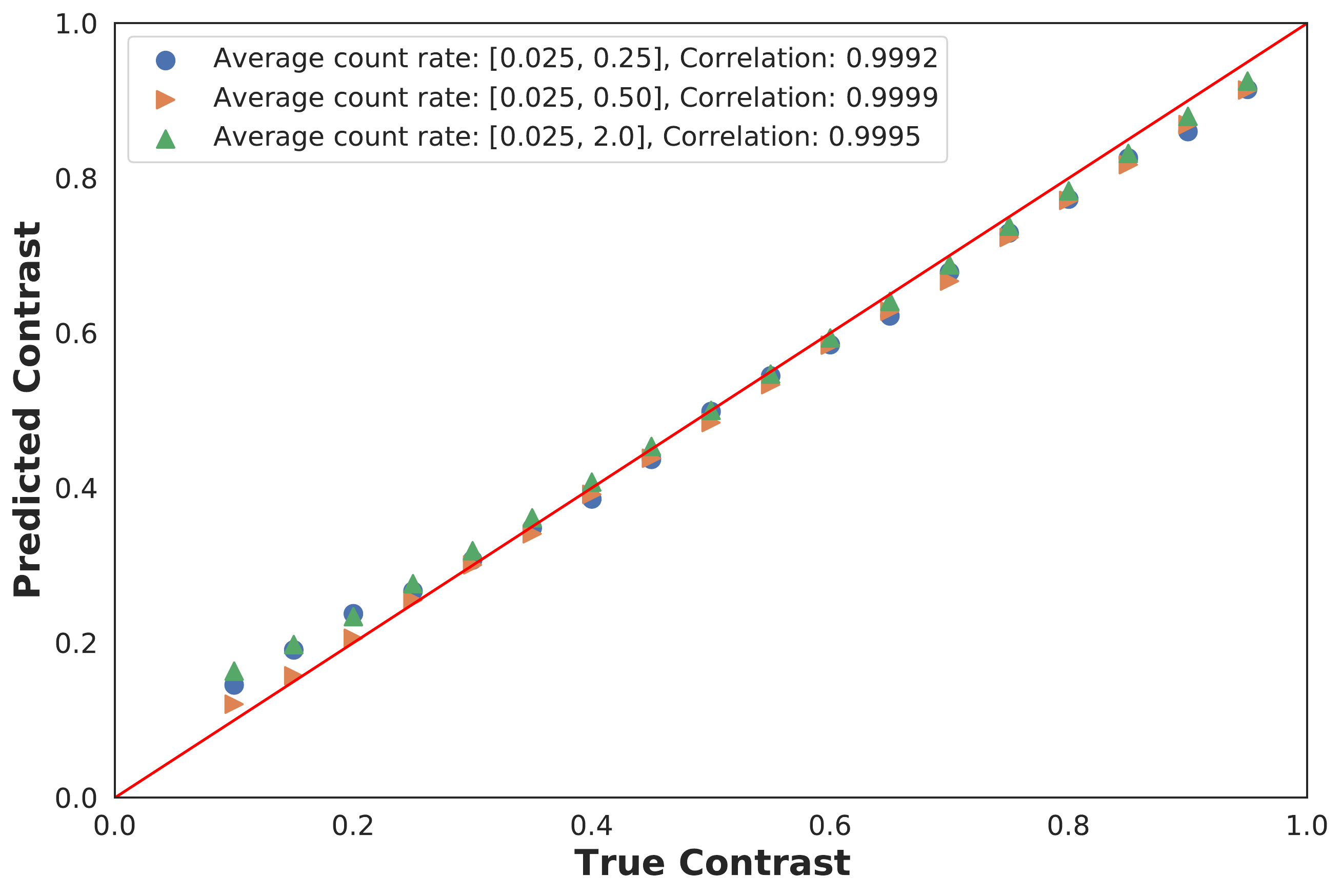}
    \caption{\label{fig:eval_metric} Correlation in contrast-contrast parity plot. Here, each data point represents the predicted contrast over 5000 frames after fitting to negative binomial statistics. The overall evaluation metric of a model is the average of the correlations for the different $\bar{k}$ ranges. In this instance, the average correlation is 0.9989.}
\end{figure}

In this analysis, we use the average correlation (across datasets with different average count-rates) of the contrast-contrast parity model as a metric for model validation. This metric can be visualized in Figure~\ref{fig:eval_metric}.

\clearpage

\section{Value of High $\bar{k}$}

\begin{figure}[h]
    \centering
    \includegraphics[width=0.6\textwidth]{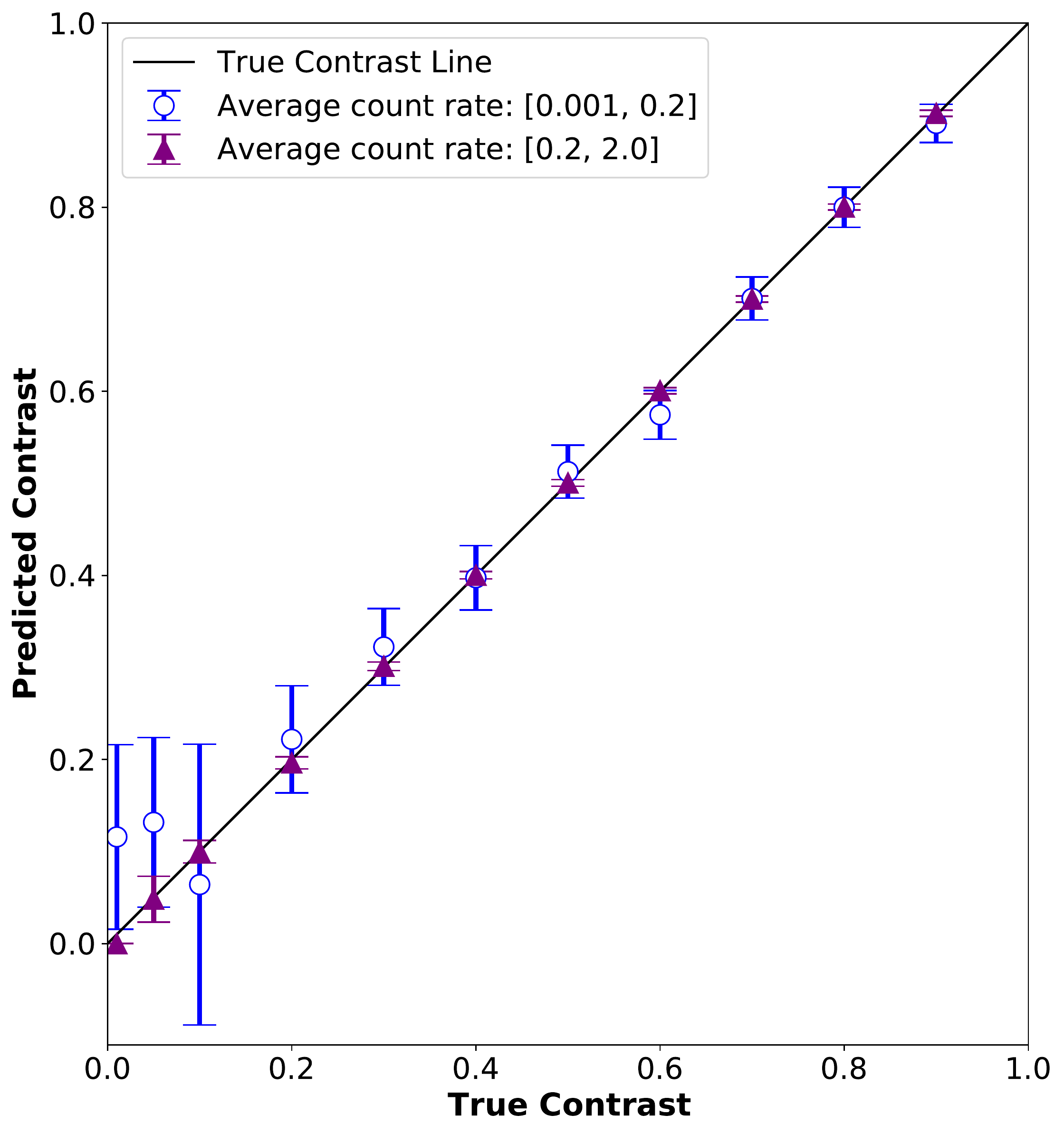}
    \caption{\label{fig:kbar_theory} Comparison of fitting for high $\bar{k}$ data vs. low $\bar{k}$ on data with perfect photon maps (no error from denoising). Evidently, the error shrinks for higher $\bar{k}$ data.}
\end{figure}

Here, we consider the value of utilizing data at higher $\bar{k}$ in the fitting process. In this scenario, the photon maps are perfect samples from the negative binomial distribution (no error from the detector image to photon map conversion). At higher $\bar{k}$, high-count photon events are more probable and contain more information about the underlying distribution relative to common events (e.g. 0 or 1 photon events). It follows naturally that incorporating higher $\bar{k}$ data allows for fitting the distribution with less data and more accuracy relative to only low $\bar{k}$ fits. In Figure~\ref{fig:kbar_theory}, we show the results of fitting the negative binomial distribution with $\bar{k}$ varying uniformly in two ranges, [0.001, 0.2] and [0.2, 2.0] for 1000 datapoints. We see that at higher $\bar{k}$, the error-bars in the contrast are much smaller relative low $\bar{k}$. 

\clearpage

\bibliography{aipsamp}

\end{document}